\documentclass[aps,amsfonts,pre,twocolumn,superscriptaddress,showpacs,longbibliography]{revtex4-1}

\usepackage[dvips]{graphicx}
\usepackage{amssymb,amsfonts,amsmath,bm}
\usepackage[usenames]{color}
\usepackage{bm}

\def\Tr{\mbox{Tr}\,}

\newcommand{\bbm}{\begin{multline}}
\newcommand{\eem}{\end{multline}}
\newcommand{\be}{\begin{equation}}
\newcommand{\ee}{\end{equation}}
\newcommand{\bea}{\begin{eqnarray}}
\newcommand{\eea}{\end{eqnarray}}
\newcommand{\p}{\partial}
\newcommand{\1}{\frac{1}{2}}

\newcommand{\cM} {\mathcal{M}}

\newcommand{\comment}[1]{}

\newcommand{\xv}{\bm{{\rm x}}}

\newcommand{\vv}{\bm{{\rm v}}}

\newcommand{\nv}{\bm{{\rm n}}}

\begin{document}

\title{Anisotropic odd viscosity via time-modulated drive}

\author{Anton Souslov}
\affiliation{Department of Physics, University of Bath, Claverton Down, Bath BA2 7AY, UK}
\affiliation{The James Franck Institute, 
The University of Chicago, Chicago, IL 60637, USA}
\author{Andrey Gromov}
\affiliation{Brown Theoretical Physics Center and Department of Physics,
Brown University, 182 Hope Street, Providence, RI 02912, USA}
\affiliation{Materials Sciences Division, Lawrence Berkeley National Laboratory and Department of Physics, University of California, Berkeley, CA 94720, USA}
\author{Vincenzo Vitelli}
\affiliation{The James Franck Institute, 
The University of Chicago, Chicago, IL 60637, USA}
\affiliation{Department of Physics, 
The University of Chicago, Chicago, IL 60637, USA}

\begin{abstract}
At equilibrium, the structure and response of ordered phases are typically determined by the 
spontaneous breaking of spatial symmetries.
Out of equilibrium, 
spatial order itself can become a dynamically emergent concept.
In this article, we show that spatially anisotropic viscous coefficients and stresses 
can be designed in a far-from-equilibrium fluid 
by applying to its constituents a time-modulated drive.
If the drive induces a rotation whose rate is slowed down when the constituents point along specific directions,
anisotropic structures and mechanical responses arise at long timescales.
We demonstrate that the viscous response of such anisotropic driven fluids
can acquire a tensorial, dissipationless component called
anisotropic odd (or Hall) viscosity. Classical fluids with internal torques can display 
additional components of the odd viscosity neglected in previous studies of quantum Hall fluids that assumed angular momentum conservation. We show that these anisotropic and angular momentum-violating odd-viscosity coefficients can change even the bulk flow of an incompressible fluid by acting as a source of vorticity. In addition, shear distortions in the shape of an inclusion result in torques.
\end{abstract}

\maketitle

In equilibrium phases of matter, large-scale structure is intricately tied to the 
spontaneous breaking of translational and rotational symmetries. 
Such equilibrium symmetry breaking occurs at phase transitions 
when the balance of entropic and energetic forces shifts.
In the broken-symmetry state, 
spatial symmetries (and conservation laws) determine the material's mechanical response.
In addition to crystallization, this overarching mechanism 
includes the transition to intermediate mesophases, such as nematic liquid crystals,
in which only rotational symmetries of the fluid are broken.

Systems far from equilibrium can display novel phases
having no equilibrium counterparts. Examples include 
active materials in which energy-consuming components can spontaneously break rotational symmetry to form a flock~\cite{Marchetti_review}, 
periodically driven Floquet systems that exhibit topological order~\cite{Kitagawa2010,Rudner2013,Rechtsman2013}, 
and quantum systems in which discrete time-translation symmetry is spontaneously broken, leading to analogues of crystals in the time domain~\cite{Keyserlingk2016,Choi2017,Zhang2017,Wilczek2012}.
In this article, we show how to use a \emph{time} modulated drive to induce \emph{spatially} anisotropic mechanical responses 
in a many-body system. The resulting non-equilibrium states differ from more conventional phases with spontaneously broken symmetry.
Unlike the more common examples of Floquet phases, we explore the dynamics on timescales
much longer than a period of the drive.
Our starting point is a hydrodynamic theory that describes an ordered liquid (e.g., a nematic)
whose orientation is prescribed purely by a strong external drive (or internal activity). 
The collective
mechanical response of these liquids with time-modulated drive emerges 
from the interplay between the dynamically induced alignment 
(which can be a single-particle effect) and the many-body interactions
between rotating constituents. Because of this coupling, temporal modulations of the drive can 
generate an anisotropic mechanical response that reflects the breaking of both
time-reversal and chiral symmetries. Such an anomalous mechanical response is captured
by time-averaged physical quantities
and does not require fine tuning of hydrodynamic coefficients or driving fields.  

\begin{figure}[t!]
	\includegraphics[angle=0,width=\columnwidth]{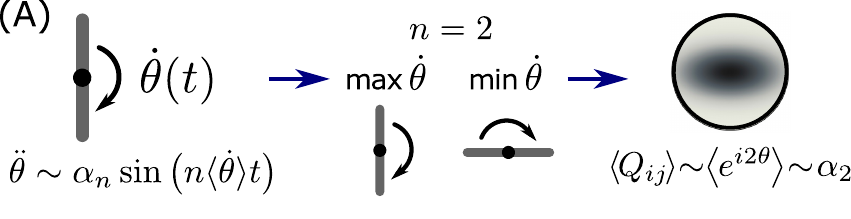}
	\caption{\label{Fig1} 
	Constructing orientational order via cyclic drive. (A) 
     Consider a fluid composed of rods (i.e., a nematic liquid crystal).
	In the model we consider, each particle rotates around its center of mass
	and the rate of rotation is modulated in time twice per cycle (left panel).
	For this case, the rods rotate fastest when oriented
	vertically and slowest when oriented horizontally (middle panel).
	On average, this means that each rod spends more time pointing
	horizontally, implying the emergence of a time-averaged nematic $\mathrm{Q}$-tensor,
	whose amplitude (i.e., the order parameter $\langle e^{i 2 \theta}\rangle$)
	is determined by the amplitude of drive modulation $\alpha$.
	The original nematic fluid
	and the rotated fluid share a $C_2$ rotational symmetry.
	However unlike equilibrium nematics, the fluid of rotating rods breaks both time-reversal and parity symmetries,
	which endows this fluid with additional mechanical response not seen in equilibrium.}
\end{figure}

The counterintuitive properties of these driven phases arise from a simple observation: in equilibrium, time-averaging and space-averaging operations must both be identical to ensemble-averaging by the ergodic theorem, whereas far from equilibrium, different averaging operations correspond to different physical quantities.
We use this principle to 
design anisotropic driven fluids with unusual mechanical properties, as illustrated in Fig.~\ref{Fig1}A.
Consider
rod-like particles for which a time-averaged nematic order parameter can be obtained by rotating the rods
with a cyclically modulated rate. When the rods point along a prescribed direction (defined by angle $\theta$), the rotation rate slows
down (corresponding to $\ddot{\theta} < 0$).
In the opposite phase of the cycle, the rods point perpendicularly to the prescribed direction 
and are sped up (with $\ddot{\theta} > 0$).
This prescribed direction defines a dynamically induced nematic order at long timescales (Fig.~\ref{Fig1}A, right panel).
The time-averaged nematic order parameter
scales with the amplitude of the modulation. 
If no modulation is present,
then the fluid appears isotropic at long timescales---this is the usual case of a chiral active fluid with a uniform rate of rotation (see right-most panels in Fig.~\ref{Fig1}B and Refs.~\cite{Sumino2012,Bonthuis2009,Furthauer2012,  Oswald2015,Riedel2005, Denk2016, Snezhko2016,Lemaire2008, Uchida2010,Yan2015}).

We consider how
the emergent fluid mechanics reflects the breaking of time-reversal, parity, and rotational symmetries
in liquids with a time-modulated drive.
We focus on a dissipationless transport coefficient called \emph{odd viscosity} (equivalently, Hall viscosity)~\cite{Avron1995, Avron1998, Read2009, read2011hall, bradlyn2012kubo, Lapa2014, Ganeshan2017}, which is represented mathematically by the anti-symmetric component of the viscosity tensor. The isotropic part of the odd viscosity tensor $\eta^o_{ijkl}$ has been studied in chiral active fluids in which each particle experiences an intrinsic torque~\cite{Banerjee2017, Liao2019}, in inviscid fluids composed of vortices~\cite{Wiegmann2014}, and in two-dimensional conductors subject to an external magnetic field~\cite{scaffidi2017hydrodynamic, delacretaz2017transport}.
This isotropic response has also been measured experimentally in colloidal chiral active fluids~\cite{Soni2019}, magnetized plasmas~\cite{Korving1966,Landau10} and graphene~\cite{berdyugin2018measuring}.
Odd viscosity arises in chiral active fluids not as a result of broken spatial symmetry, but
rather as a result of broken time-reversal symmetry, $T$.
For a simple fluid with $T$-symmetry, the Onsager reciprocal relation (valid at equilibrium) dictates that $\eta_{ijkl} = \eta_{klij}$. Without $T$-symmetry, an extra component $\eta^o_{ijkl}$ ($= - \eta^o_{klij}$) can enter the viscosity tensor with the property that both $T$ and parity operator $P$ change the sign of $\eta^o_{ijkl}$.

To see how anisotropic terms in the odd viscosity tensor
affect the fluid mechanics, 
we follow the approach developed in Ref.~\cite{Scheibner2019} and express the stress  $\sigma_{ij}$ ($ = \eta^o_{ijkl} \p_k v_l\,$) in terms of four independent components: (1) anti-symmetric stress ($\epsilon_{ij}\sigma_{ij}$), (2) isotropic pressure ($\Tr{\sigma}$), and (3,4) two Pauli matrices $\sigma^{x,z}_{ij}$ corresponding to the shear stresses at 45$^\circ$ with respect to each other. Similarly we decompose the strain rates $\partial_k v_l$ in terms of (1) vorticity 
$\omega$ ($=\epsilon_{kl}\partial_k v_l$), (2) compression $\nabla \cdot \vv$, and (3,4) two shear-strain rates. 
In the visual notation of Ref.~\cite{Scheibner2019}, the anti-symmetric component of the viscosity tensor takes the schematic form: 
\begin{equation}
\includegraphics{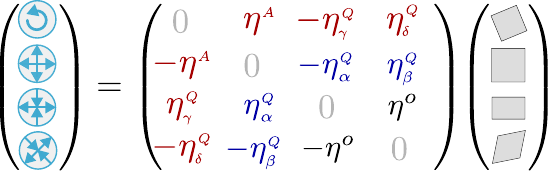}
\label{eq:as}
\end{equation}
The six independent components can be split into two groups: the two isotropic components $\eta^{o}$ and $\eta^{A}$, and the four components that transform under rotation $\eta^Q_{\alpha}$, $\eta^Q_{\beta}$, $\eta^Q_{\gamma}$, and $\eta^Q_{\delta}$.
The usual isotropic odd viscosity $\eta^{o}$ couples the two shear components corresponding to $\sigma^{x}_{ij}$ and $\sigma^{z}_{ij}$ in a chiral fashion. By contrast, the $\eta^{A}$ component corresponds to local torques due to fluid compression and explicitly violates the conservation of angular momentum. Similarly,
the anisotropic components $\eta^Q_{\gamma}$ and $\eta^Q_{\delta}$ generate antisymmetric stress and only appear in fluids that violate the conservation of angular momentum, whereas $\eta^Q_{\alpha}$ and $\eta^Q_{\beta}$ are their angular-momentum-conserving counterparts. Whereras quantum Hall fluids (including anisotropic ones) conserve angular momentum and have $\eta^{A} = 0$, chiral active fluids do exhibit a nonzero $\eta^{A}$ even in the isotropic case.  
The anisotropic components can be split into two pairs:  (1) $\eta^{Q}_{\alpha,\beta}$ leads to pressure (i.e., isotropic stress) due to shear and vice versa, in a direction-dependent way, and (2) $\eta^{Q}_{\gamma,\delta}$ leads to torque (i.e., antisymmetric stress) due to shear and vice versa. Under a $45^\circ$ coordinate rotation, $\eta^{Q}_{\alpha}$  transforms into $\eta^{Q}_{\beta}$, $\eta^{Q}_{\gamma}$ transforms into $\eta^{Q}_{\delta}$, and the (squared) amplitudes  $(\eta^Q)^2 \equiv (\eta^{Q}_{\alpha})^2+(\eta^{Q}_{\beta})^2$ and $(\eta^K)^2 \equiv (\eta^{Q}_{\gamma})^2+(\eta^{Q}_{\delta})^2$ remain invariant.

Phenomenologically, we show how to distinguish anisotropic odd viscosity from the previously investigated isotropic part. For an incompressible fluid with conserved angular momentum, isotropic odd viscosity (characterized by $\eta^{o}$) affects the pressure but not the fluid flow profile~\cite{Avron1998,Lapa2014,Ganeshan2017}. Without angular momentum conservation, an isotropic incompressible fluid still exhibits no signature of the extra odd viscosity $\eta^{A}$. In summary, isotropic odd viscosity cannot be measured from an incompressible flow profile~\cite{Avron1998,Banerjee2017}. We show that by contrast, the anisotropic-odd-viscosity components $\eta^{Q}_{\gamma,\delta}$ explicitly enter the equation of motion for the vorticity, $\omega$, of an incompressible fluid through the symmetric traceless matrices ${\mathcal M}_1 \equiv \eta^Q_{\gamma} \sigma^x + \eta^Q_{\delta} \sigma^z$ (which is proportional to the $Q$-tensor and where $\sigma^{x,z}$ are the Pauli matrices) and ${\mathcal M}_1^* \equiv \eta^Q_{\delta} \sigma^x - \eta^Q_{\gamma} \sigma^z$ (i.e., $\cM_1$ rotated by $\pi/4$),
\begin{align}
\label{eq:omm}
&\rho D_t \omega = \eta \nabla^2 \omega  - (\nabla \cdot \cM_1 \cdot \nabla) \omega + \nabla^2 [\nabla \cdot (\cM_1^* \cdot \vv)],
\end{align}
where $\rho$ is the density and $\eta$ is the dissipative shear viscosity. The last terms represent torques induced by the shear components of the strain rates due to the anisotropic odd viscosities $\eta^{Q}_{\gamma,\delta}$.
For a parity-violating fluid with conserved angular momentum, anisotropic odd viscosity can still be measured via torques on a shape-changing inclusion.
References~\cite{Lapa2014, Ganeshan2017} show that isotropic odd viscosity results in torques on an inclusion proportional to the rate of change in area.
Here we show that the anisotropic odd viscosity components $\eta^Q_{\alpha,\beta}$ capture an additional effect corresponding to torques that result from the change in the shape of an inclusion at fixed area, i.e., from the shear distortions of the inclusion's boundary (see Fig.~\ref{Fig2}). 

\section{\label{sec:struc} Nematic phases from time-averaging}
In this Section, we derive a coarse-grained description
for the structure of a fluid composed of rapidly rotating anisotropic objects. 
Define the director to be 
$\hat{\mathbf{n}}(t) = [\cos \theta(t), \sin \theta(t) ]$
and modulate the orientational dynamics of the rods via the angle 
$\theta(t)$:
\begin{equation}
\theta(t) = \Omega t - \alpha \sin (2 \Omega t + \delta), \label{eq:theta}
\end{equation}
where $\alpha$ is the modulation amplitude, $\Omega = \langle \dot{\theta}(t) \rangle$ is the average rotation rate, $\delta$ is the rotation phase, and the averaging is over a period of rotation from $t = 0$ to 
${t = 2 \pi/\Omega}$.

In the context of equilibrium spontaneous symmetry breaking,
the constituent shape determines mesophase order.
For example, at high density or low temperature, rod-shaped constituents can form nematic (two-fold rotationally symmetric) phases. 
By contrast, in our case, anisotropic responses and structure emerge from dynamics.
In order to characterize such structure on long timescales, we average
over the fast timescale of a single rotation period.
We formally define this time-averaging via the integral 
\begin{equation}
\langle \chi(t) \rangle \equiv \frac{\Omega}{2 \pi} \int_0^{2 \pi/\Omega} \! dt \,\, \chi(t)
\label{eq:av}
\end{equation}
for an arbitrary periodic function $\chi(t)$. 
For example, substituting Eq.~(\ref{eq:theta}), with $\delta = 0$, into the orientational order parameter $e^{i 2 \theta}$
and evaluating the average using Eq.~(\ref{eq:av}), we find
\begin{equation}
\langle e^{i 2 \theta(t)} \rangle =  J_1(2 \alpha) \approx \alpha + O(\alpha^3), \label{eq:op}
\end{equation}
where $J_1(x)$ is a Bessel function of the first kind~\footnote{This expression can be obtained using the definition $
J_1(x) \equiv \frac{1}{2 \pi}\int_{-\pi}^{\pi}e^{i(\tau - x \sin \tau)} d\tau $.}.
This order parameter connects the modulation defined by Eq.~(\ref{eq:theta}) to time-averaged orientational order with $2 \pi$ rotational symmetry.
In the isotropic case, the system becomes a fluid composed
of objects rotating at a constant rate~\cite{Sumino2012,Bonthuis2009,Furthauer2012,Oswald2015,Riedel2005, Denk2016, Snezhko2016,Lemaire2008, Uchida2010,Yan2015}.
The mechanics of matter composed of such chiral active building blocks is crucial for biological function~\cite{Drescher2009, Petroff2015,Nonaka2002,Guirao2010,Button2012,Brumley2015,Kirchhoff2005,Kaiser2013,Lenz2003}
and synthetic materials design~\cite{Tabe2003, Maggi2015, Nguyen2014,Sabrina2015,Spellings2015}.
One exotic feature in the mechanics of these fluids are local torques due to antisymmetric components of the stress tensor~\cite{Dahler1961, Condiff1964, Tsai2005}.

The order parameter captures the appearance of nematic anisotropy in a fluid with a cyclically modulated drive. Rotations of time-averaged order are captured by modulation phase $\delta$ that enter the nematic $\mathrm{Q}$-tensor.
(The order parameter $S \equiv |\langle e^{i 2 \theta(t)} \rangle|$ does not depend on rotations by $\delta$.)
For a fluid with nematic symmetry, the time-averaged $\mathrm{Q}$-tensor is defined by $\langle Q_{ij} \rangle  \equiv 2 (\langle n_i n_j \rangle - \langle n_i n_j \rangle_{\alpha = 0})$, where  $\langle n_i n_j \rangle_{\alpha = 0} = \delta_{ij}$ is the average in the isotropic case ($\delta_{ij}$ is the Kronecker-$\delta$).
Using Eq.~(\ref{eq:theta}), we find:
\begin{align}
\label{eq:qdef}
\langle Q_{ij} \rangle & = \frac{S}{2} \begin{bmatrix}
    \cos2\delta & \sin2\delta \\
    \sin2\delta & - \cos2\delta
\end{bmatrix}.
\end{align}
In this time-averaged sense, the fluid is not an ordinary nematic, which would have a spontaneously broken symmetry and long, slow variations in $Q_{ij}(\xv,t)$ over time and space. Instead, in the driven fluid such fluctuations are suppressed because rotational symmetry is explicitly broken by the drive. $Q_{ij}$ is prescribed and constant in both time and space.

For this nematic fluid, the naive time-average of the director $\hat{\nv}$ is zero by symmetry: $\langle \hat{\nv} \rangle = 0$. Nevertheless, a time-averaged director $\hat{n}^a$ can be defined from the time-averaged $\mathrm{Q}$: $\langle Q_{ij} \rangle =\langle e^{i 2 \theta(t)} \rangle [\hat{n}^a_i \hat{n}^a_j - \delta_{ij}/2]$. This quantity is defined by the phase $\delta$, 
$\hat{n}^a = (\cos \delta, \sin \delta)$.
The two parameters $\alpha$ and $\delta$ determine, respectively, the magnitude and orientation of the time-averaged order in the emergent nematic fluid (as does the equivalent description using the $\mathrm{Q}$-tensor). 

\section{\label{sec:model2} Anisotropic odd viscosity}
A fluid with orientational order has a direction-dependent mechanical response.
Here, we ask ``does time-modulated drive lead to a response, for example in the viscosity tensor,
which is not possible in equilibrium?''
To probe such viscosities, we consider timescales for which $\dot{\theta}$ is fast and the strain rates $\nabla_i v_j$ are slow.
In our analysis, we begin with a coarse-grained description of an equilibrium nematic liquid crystal
and add drive. Such a description is appropriate if $\dot{\theta}$ is slow compared to the microscopic collision processes between the fluid particles, allowing us to keep only the lowest-order terms in $\dot{\theta}$.
In our description, the fast director is averaged over a rotational period,
and only the slow velocity field remains (see Fig.~\ref{Fig1}).

For general two-dimensional fluids that conserve angular momentum, the odd viscosity encoded in the tensor $\eta^o_{ijkl}$ ($= - \eta^o_{klij}$) has three independent components $\eta^Q_{\alpha,\beta}$ and $\eta^o$~\cite{Avron1998}.
Because the driven rotation is a clear source and sink for angular momentum in the overdamped fluid that we consider, odd viscosity includes the three extra components $\eta^Q_{\gamma,\delta}$ and $\eta^A$.
Whereas the components $\eta^{o,A}$ are isotropic, the $\eta^Q_{\alpha,\beta,\gamma,\delta}$ rotate like the components of the $Q$-tensor for a nematic liquid crystal. Therefore, for a fluids with three-fold rotational symmetry (or higher), only the two isotropic components $\eta^{o,A}$ will remain~\cite{Avron1998}.
Note that for any odd viscosity tensor $\eta^o_{ijkl}$, the resulting stress $\eta^o_{ijkl} v_{kl}$ is dissipationless. This can be evaluated from the rate $\partial_t s$ of entropy production, $\partial_t s \approx \sum_{ijkl}\eta^{o}_{ijkl} v_{ij}v_{kl} = 0$ using the anti-symmetry of $\eta^o_{ijkl}$. Odd viscosity may be a useful tool in the study of parity-broken quantum systems such as quantum Hall states, Chern insulators, and topological superconductors~\cite{read2011hall, abanov2014electromagnetic, bradlyn2015low, gromov2015framing, gromov2016boundary}, because this anomalous response can be used to identify topological phases of matter.

For two-dimensional quantum fluids,
an anisotropic generalization of odd viscosity has recently been proposed in Refs.~\cite{Haldane2009, Haldane2015, Gromov2017, lapa2018hall}.
In these cases, the fluid has inversion symmetry as well as angular momentum conservation,
and the full information about odd viscosity is encoded into a symmetric rank-2 tensor $\eta^{o}_{ij}$:
\be
\eta^o_{ij} = \eta^o \delta_{ij} + \eta^Q_{\alpha} \sigma^x_{ij} +  \eta^Q_{\beta} \sigma^z_{ij},
\label{eq:oddten}
\ee
where the traceless part of $\eta^o_{ij}$ is the symmetric matrix $\eta^Q_{\alpha} \sigma^x + \eta^Q_{\beta} \sigma^z$.
As an example, if the nematic director aligns with the x-axis, then $\delta = 0$. Physically, this means that only the horizontal pure shear leads to either a torque or a pressure change. 
Isotropic odd viscosity $\eta^o$ has been observed in magnetized plasmas~\cite{Korving1966, Landau10}, whereas nematic components of odd viscosity have not yet been realized in any experimental context.
In order to estimate anisotropic odd viscosity in chiral active fluids, 
we begin with an anisotropic classical fluid with overdamped orientational dynamics, i.e., a nematic liquid crystal~\cite{deGennes1995,Chandrasekhar1992}.
Typical nematics are composed of anisotropic, rod-like constituents (called nematogens)
on molecular or colloidal scales. 
When the rods align with their neighbors,
they carry no angular momentum or inertia.
Vibrated rods can order into a nematic pattern
as a nonequilibrium example of a system with liquid-crystalline order~\cite{Galanis2006}.
Nematogens
can transition between a disordered state at high temperature (or low density)
and an aligned state at low temperature (or high density).
In the nematic state,
the rods tend to all point in the same direction,
and the mechanical response varies relative to this alignment.
The Leslie-Ericksen coefficients characterize the linear response of the fluid stress
to either the strain rate or the rotation rate of the nematic director.  

We now consider the nonlinear generalization of the Leslie-Ericksen stress, to lowest orders in nonlinearities~\cite{Moritz1976}
(see Supporting Information for full expression).
After averaging over the fast dynamics of the nematic director, the terms linear in strain rate $A_{ij}$ contribute to the viscous components of the stress tensor.
However, terms even in $\dot{\hat{\nv}}$ (i.e., order $(\dot{\hat{\nv}})^{2p}$ for integer $p$, including $p=0$, which are those independent of $\dot{\hat{\nv}}$) do not break time-reversal symmetry and cannot contribute to odd viscosity. 
We focus on those terms that contribute to the odd viscosity tensor, which therefore must be odd in 
$\dot{\hat{\nv}}$ ($\dot{n}_i = - \dot{\theta} \epsilon_{ij} n_j$, where
$\epsilon_{ij}$ is the two-dimensional Levi-Civita symbol defined via $\epsilon_{xy} = - \epsilon_{yx} = 1$ and $\epsilon_{xx} = \epsilon_{yy} = 0$) and linear in $A_{kl}$.
For positive integers $\beta$ ($ = 1,2,3,\ldots$), these terms, of order $\dot{\theta}^{2 \beta - 1}$, are~\cite{Moritz1976}
\begin{align}
& \sigma^{EL,\beta}_{ij} = \dot{\theta}^{2 \beta - 2 } \big[ \xi^\beta_{10} n_p A_{ip} \dot{n}_j 
+ \xi^\beta_{11} n_p A_{jp} \dot{n}_i + \xi^\beta_{12}  n_i A_{jp} \dot{n}_p  \nonumber \\
 & + \xi^\beta_{14}  n_j A_{ip}  \dot{n}_p
+ \xi^\beta_{16} n_i n_p n_q A_{pq} \dot{n}_j + \xi^\beta_{17} n_j n_p n_q A_{pq} \dot{n}_i \big]\,.
\label{eq:el1}
\end{align}
We focus on the stress components $\sigma^{EL,1}$ and $\sigma^{EL,2}$,
which have similar forms, but different orders of $\dot{\theta}$ and, in general, different sets of coefficients
$\{\xi^\beta_\kappa\}$.
The local forces $\rho_0 \partial_t \vv$ are calculated using gradients of the time-averaged stress, resulting in the equation for the flow $\vv$:
$\rho_0 \partial_t v_i = \nabla_j \langle \sigma^{EL}_{ij}\rangle$, where $\rho_0$ is the fluid density.
For modulations with $n = 2$, we obtain the following expression for isotropic odd viscosity:
\begin{equation}
\label{eq:etao1}
\eta^o = - \frac{\Omega}{8} \xi^1_L - \frac{\Omega^3}{8} (1 + 2 \alpha^2 ) \xi^2_L + O(\Omega^5),
\end{equation}
where $\xi^\beta_L \equiv 2 [\xi^\beta_{10} + \xi^\beta_{11} - \xi^\beta_{12} - \xi^\beta_{14}] 
+ \xi^\beta_{16} +\xi^\beta_{17}$ is a linear combination of the $\xi^\beta_\kappa$ coefficients.
The first right-hand-side term in Eq.~[\ref{eq:etao1}] comes from the lowest-order nonlinearities in the equilibrium fluid stress, whereas
the higher-order term involves higher-order nonlinearities and will in general be subdominant.
Despite constraints (stemming from stability at equilibrium) on the signs of $\xi^\beta_i$, the resulting expression~(\ref{eq:etao1}) for $\eta^o$ can change sign either 
via reversal of the spinning rate $\Omega$ or by changing the relative magnitudes of $\xi^\beta_\kappa$ that enter Eq.~(\ref{eq:etao1}) with different signs.

To analyze the tensorial (angular-momentum conserving) components of the odd viscosity tensor $\eta^o_{ijkl}$,
we calculate the rank-2 odd viscosity tensor $\eta^o_{ij}$ using~\cite{Haldane2009, Haldane2015,Gromov2017}
$\eta^o_{ij} = (\delta_{ni} \delta_{kj} \epsilon_{ml} + \delta_{mi} \delta_{lj} \epsilon_{nk}) \eta^o_{nmkl}/4$.
From $\langle \sigma^{EL,2}_{ij} \rangle$, we find
\begin{align}
\eta^Q & = \frac{\alpha \Omega^3}{4} (\xi^2_{16} + \xi^2_{17}) + O(\Omega^5),
\end{align}
where again $\eta^Q$ is defined via $(\eta^Q)^2 \equiv (\eta^{Q}_{\alpha})^2+(\eta^{Q}_{\beta})^2$.
Because effects of modulated drive enter via terms of the stress $\sigma_{ij}$ higher-order in the rotation rate $\dot{\theta}$, $\eta^Q$ scales as $\Omega^3$ in contrast to $\eta^o$, which scales as $\Omega$.
If $\alpha \rightarrow 0$, the driven fluid loses anisotropy and the nematic odd viscosity
$\eta^Q$ vanishes.

In addition to the components of the odd viscosity tensor that conserve angular momentum, the chiral active fluid also includes the components $\eta^Q_{\gamma,\delta}$ and $\eta^A$ that couple explicitly to the antisymmetric component of the stress and which therefore correspond to induced microscopic torques. These out-of-equilibrium responses differentiate the far-from-equilibrium fluid from, for example, quantum Hall fluids that break time-reversal symmetry at equilibrium due to an applied magnetic field and which instead do conserved angular momentum.
From the averaging procedure, these extra responses can be read off as 
\begin{align}
    &\eta^A = \frac{\Omega}{4} (- \xi^1_{9} + \xi^1_{10} - \xi^1_{14}) + O(\Omega^3)\\
    &\eta^{K}  = \frac{\alpha \Omega^3}{4} (2 \xi^2_{11} + 2 \xi^2_{12} - \xi^2_{16} + \xi^2_{17})  + O(\Omega^5)
\end{align} 
to lowest orders in $\Omega$, where $\eta^K$ is defined via $(\eta^K)^2 \equiv (\eta^{Q}_{\gamma})^2+(\eta^{Q}_{\delta})^2$.

In many contexts, odd viscosity goes hand in hand with inertia.
In vortex fluids, the vortex circulation
encodes both fluid inertia and odd viscosity~\cite{Wiegmann2014}. 
For chiral active fluids in which collisions conserve angular momentum, 
a simple argument gives the value of odd viscosity: if an inclusion
changes its area, the torque on the inclusion is given by the rate of change in area
times the odd viscosity or, equivalently, by the expelled angular momentum.
As a result, odd viscosity is given by half of the angular momentum density~\cite{Ganeshan2017,Banerjee2017}.

For fluid phenomena at the smallest scales, dissipation dominates over inertia.
In this limit, chiral active fluids composed of colloidal particles have the broken-$T$ symmetry
necessary for odd viscosity to arise. However, the arguments based on angular momentum
cannot give an accurate estimate of the value of odd viscosity because momentum plays no role in the mechanics.
Instead, in the dissipative, overdamped model that we propose, isotropic odd viscosity $\eta^o$
arises from the lowest-order nonlinear coupling between director rotation and fluid strain rate.
Furthermore, in a fluid with broken time-reversal, parity, and rotational symmetries,
higher-order nonlinear couplings lead to an anisotropic components $\eta^{Q,K}$ 
of the odd viscosity tensor.

\begin{figure}[t!]
	\includegraphics[angle=0]{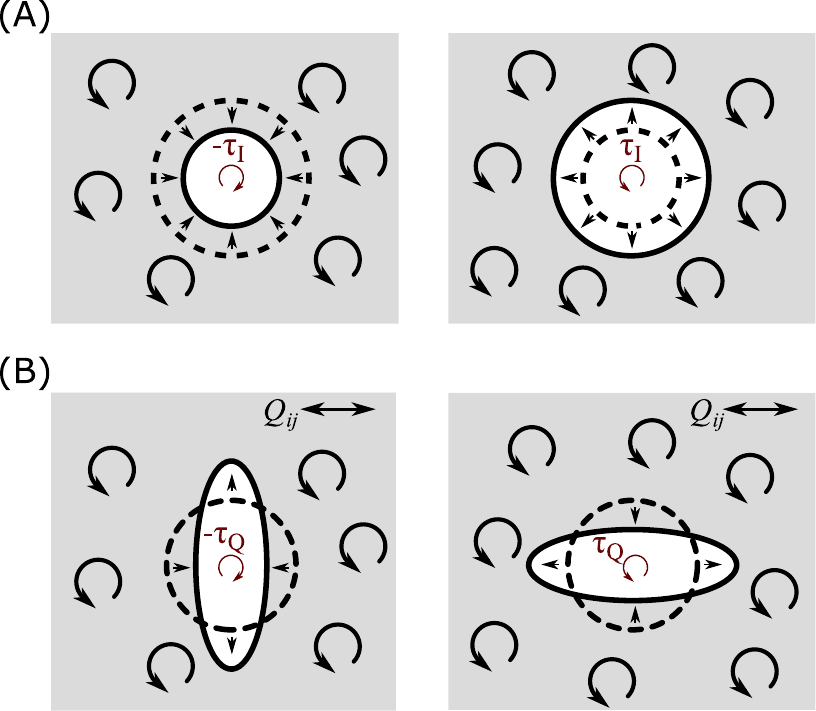}
	\caption{
	Schematics of the physics of tensorial odd viscosity.
     (a) The response characteristic of isotropic odd viscosity, 
     corresponsing to $\eta^o = \mathrm{Tr}(\eta^o_{ij}) /2$:
     for an object with time-varying area $a(t)$, isotropic odd viscosity
     is related to the ratio of torque $\tau_I$ to areal rate of change $\dot{a}$: 
     $\eta^o = \tau_I/(2\dot{a})$~\cite{Lapa2014,Ganeshan2017}.
     For a given fluid chirality (in this case, $\eta^o > 0$), 
     the torque changes sign depending on whether the object is 
     contracting ($\dot{a} < 0$ and $\tau_I < 0$, left) or expanding ($\dot{a} > 0$ and $\tau_I > 0$, right).
     (b) If the areal rate of change is zero, but the shape is sheared,
     then the torque $\tau_Q$ is given by the anisotropic component of the odd viscosity tensor.
    This nematic odd viscosity has two independent components captured by the traceless symmetric tensor
   $Q_{ij}$ [$= S (n_i n_j - \delta_{ij}/2)$],
    which control the amplitude and shear-angle-dependence of the resulting torque.
     Specifically, this torque depends on the angle of the shear relative
     to the director $n_i$ and is proportional to the (signed) shear rate.
     For example, for a sheared circle, a rotation of the shear by $\pi/2$
     is equivalent to a shear of opposite sign,
     and therefore corresponds to a torque $\tau_Q$ of the opposite sign (right). 
     The orientation at angle $\pi/4$ at which the shear is diagonal
     corresponds to zero torque.}
	\label{Fig2}
\end{figure}

\section{\label{sec:torque1}Equation of motion with anisotropic odd viscosity}
In this section, we show the consequences of tensorial odd viscosity on fluid flow.
Using the Helmholtz decomposition in two dimensions, the fluid flow can be 
expressed in terms of the compression rate $\nabla \cdot \vv$ and the vorticity $\nabla \times \vv$. 
To derive the equation of motion for vorticity, we follow the usual route by taking the curl of the velocity equation.
This simplifies the equation by removing the gradient terms due to isotropic stress (because $\epsilon_{ij} \partial_i \partial_j \sigma_{kk} = 0$). Without any odd viscosity contributions, the equation of motion would become the two-dimensional vorticity-diffusion equation.
We find that whereas isotropic odd viscosity contributes only compression-rate-dependent terms, anisotropic odd viscosity changes the vorticity profile even for an incompressible fluid~\cite{Banerjee2017}. We do so by substituting the expression for the stress $\sigma_{ij} = \eta_{ijkl} v_{kl}$ into the velocity equation $\rho D_t v_j = \partial_i \sigma_{ij}$. We begin with the full anti-symmetric viscosity tensor $\eta^o_{ijkl}$ from Eq.~(\ref{eq:as}) and, for brevity, only the isotropic shear viscosity $\eta$ from the symmetric, dissipative viscosity (see Appendix~\ref{sec:apdis} for a detailed discussion of the anisotropic dissipative viscosity tensor.) 
Taking the curl, we arrive at the (pseudo-scalar) vorticity equation (see Appendix~\ref{sec:derom} for details):
\begin{align}
&\rho D_t \omega  =\eta \nabla^2 \omega - (\nabla \cdot \cM_1 \cdot \nabla) \omega + \nabla^2 [\nabla \cdot (\cM_1^* \cdot \vv)] \nonumber
 \\&
+ (\eta^o + \eta^A) \nabla^2 (\nabla \cdot \vv) 
-  (\nabla \cdot {\mathcal M}_2 \cdot \nabla) (\nabla \cdot \vv) \label{eq:omm2},
\end{align}
where $D_t$ is the convective derivative, and
\begin{align}
\label{eq:mdef}
\cM_1 & \equiv \eta^Q_{\gamma} \sigma^x + \eta^Q_{\delta} \sigma^z, \\
\cM_2 & \equiv \eta^Q_{\alpha} \sigma^x + \eta^Q_{\beta} \sigma^z \nonumber
\end{align}
and $\cM_1^* \equiv \eta^Q_{\delta} \sigma^x - \eta^Q_{\gamma} \sigma^z$ (i.e., $\cM_1$ rotated by $\pi/4$).
For incompressible flow, $\nabla \cdot \vv = 0$, and the last two terms in Eq.~(\ref{eq:omm2}) proportional to the odd viscosity components $\eta^{o}$, $\eta^{A}$, and ${\mathcal M}_2$ all vanish~\cite{Avron1998}. This reduces Eq.~(\ref{eq:omm2}) to Eq.~(\ref{eq:omm}). This feature distinguishes components of anisotropic odd viscosity ${\mathcal M}_1$ (and $\eta^{K}$)
from both isotropic odd viscosities $\eta^{o}$ and $\eta^{A}$: ${\mathcal M}_1$ can be measured directly from the flow of an incompressible fluid in the bulk. The expression $\nabla \cdot ({\mathcal M}^*_1 \cdot \vv)$ can be interpreted as a shear-strain rate associated with $\vv$ (because $\mathrm{Q}$ and ${\mathcal M}_{1,2}$, like shear transformations, are all symmetric and traceless). Alternatively, we can rewrite the last term in Eq.~(\ref{eq:omm}) using the nematic director rotated by $\pi/4$, which we call $\hat{\mathbf{m}}$, finding the term proportional to $\nabla^2 [(\hat{\mathbf{m}} \cdot \nabla) (\hat{\mathbf{m}} \cdot \vv)] $, where we used $\nabla \cdot \vv = 0$. This form demonstrates that anisotropic odd viscosity induces torques due to (the Laplacian of) gradients that are rotated by $\pi/4$ relative to the nematic director of the velocity component along the same direction.

A further simplification to these expressions can arise in fluids with nematic symmetry. In that case, we expect both ${\mathcal M}_1$ and ${\mathcal M}_2$ to be proportional to the nematic $Q$-tensor, which implies that the angle $\delta$ defined in Eq.~(\ref{eq:qdef}) is the same for the two tensors ${\mathcal M}_{1,2}$. This implies a relation between components $\eta^Q_{\alpha,\beta,\gamma,\delta}$ that reduces the number of independent anisotropic viscosities from four to three. This relation between the four anisotropic odd viscosities is expected to hold for a wide range of models of anisotropic fluids with odd viscosity and without angular momentum conservation, including the one we consider in this work.

\section{\label{sec:torque2}Torques on an inclusion}
Whereas the anisotropic component, $\eta^K$, can be measured directly from the flow of an incompressible fluid, the other tensorial odd viscosity, $\eta^Q$, requires the measurement of forces.
Below, we show how tensorial odd viscosity $\eta^Q$ determines the mechanical forces that the fluid exerts on immersed objects.
For simplicity, consider the case in which $\eta^A = \eta^{K} = 0$. This case also applies to the quantum Hall fluid, because the consevation of total angular momentum is preserved. We find that such a fluid exerts torques due to the shape change of the object.
We calculate the torque on a shape-changing object
by integrating the local force over the object's boundary.
We focus on expressions that apply to both inertial and overdamped fluids
by only considering the instantaneous forces $f_j$ on the boundary element of the object (and not the flow away from the boundary). These forces are determined from the instantaneous velocity $\vv$ via the fluid stress tensor $\sigma_{ij}$:
\be
f_j=m_i \sigma_{ij}\, ,
\ee
where $m_i$ is the normal to the boundary at that point. 
We then substitute into the odd-viscosity stress $\sigma_{ij}$ ($ = \eta^o_{ijkl} \p_k v_l\,$) the (general)  expression~\cite{Haldane2009, Haldane2015,Gromov2017}
\be
\eta^o_{ijkl} = \frac{1}{2} \left( \epsilon_{ik} \eta^o_{jl} + \epsilon_{jk} \eta^o_{il}+ \epsilon_{il} \eta^o_{jk}+ \epsilon_{jl} \eta^o_{ik}\, \right).
\label{eq:eo}
\ee
The force on an element of the boundary of an inclusion is given by
\be
f_j = \frac{1}{2} \left( m_k \eta^o_{jl} \p_k^* v_l + m_i \eta^o_{il}\p_j^*v_l + m_i \eta^o_{kj}\p_kv^*_i+m_i\eta^o_{ik}\p_kv^*_j \right)
\ee
where we have used the notation $v^*_i \equiv \epsilon_{ij} v_j$.

The total torque $\tau$ on a compact inclusion is given by the integral of the local torque ${\cal T}$ acting on an infinitesimal boundary element, $\tau=\oint {\cal T}(s) ds$, where $s$ is an arc-length parameterization of the boundary. The local torque is given by the standard expression ${\cal T} = \epsilon_{ij}x_i f_j = \vec{x} \times \vec{f}$.
For example, in the isotropic case $\eta^o_{ij} = \eta^o \delta_{ij}$, one obtains the relation derived in 
Refs.~\cite{Lapa2014, Ganeshan2017}:
\be
\tau_I = 2 \oint N_i \eta^o_{ij}v_j = 2 \eta^o\oint v_{N}=2 \eta^o\dot{a},
\ee
where $\dot{a}$ the rate of change of area for the inclusion and $N_i$ is the normal to the inclusion boundary.
Substituting Eq.~(\ref{eq:oddten}) into the expression for the 
integrand of the torque, we find
\begin{align}
N_i \eta^o_{ij}v_j & = \eta^o v_N + \eta^Q_{\alpha}  \sigma^{x}_{ij} N_i v_j + \eta^Q_{\beta} \sigma^{z}_{ij} N_i v_j.
\end{align}
Thus, the contribution $\tau_Q$ to the torque due to nematicity is
\begin{align}
\tau_Q & =  2 \eta^Q_{\alpha} \oint \sigma^{x}_{ij} N_i v_j + 2 \eta^Q_{\beta} \oint\sigma^{z}_{ij} N_i v_j. 
\end{align}

For a circle of radius $r_0$ at the origin, a deformation with a zero change in area and a nonzero shear rate (applied affinely, i.e., uniformly across the entire shape)
is captured by the second angular harmonic of the velocity field,
\be
f_2(\gamma) = \int d\theta \cos (2\theta-2\gamma) v_N(\theta),
\label{eq:seg}
\ee
where $v_N(\theta) = \vv(r = r_0, \theta) \cdot \hat{\mathbf{N}}$
is the normal (i.e., radial) displacement of the circle's boundary (see Fig.~\ref{Fig2}). 
The parameter $\gamma$ sets the angle of the applied shear.
To better intuit  Eq.~(\ref{eq:seg}), the angular dependence can be contrasted with areal deformation, which corresponds to the zeroth angular harmonic,
$\int d\theta v_N(\theta)$ ($=\dot{a}$), and a net translation at fixed shape, which corresponds
to the first harmonic, $\int d\theta [\cos\theta,\sin\theta]v_N(\theta)$ ($=[v_x,v_y]$).
To evaluate $\tau_Q$, we use the relation $Q_{ij} N_i N_j = \frac{S}{2}\cos(2 \theta - 2 \delta)$
and assume that $v_i = v_N N_i$, i.e., the velocity is normal to the boundary. We then find 
\begin{equation}
\tau_Q = 2 \eta^Q \oint Q_{ij} N_i N_j v_N =  \eta^Q f_2(\delta),
\end{equation}
where we used Eq.~(\ref{eq:seg}).
The torque magnitude is set by the nematic part of the odd viscosity tensor, $\eta^{Q}$, and the angular dependence is set by the nematic director angle $\delta$.
The $\eta^{Q}$ component of the nematic odd viscosity can be measured from the ratio $\tau_Q/f_2(\delta)$, i.e., measuring the torque $\tau_Q$ due to a shear rate $f_2(\delta)$ in a direction along which $f_2(\delta) \ne 0$ (see Fig.~\ref{Fig2}).
Note that $\eta^{Q}_{\alpha,\beta}$ are two independent components
of the odd viscosity tensor: these could be defined, for example,
in terms of the torque amplitude and the direction of largest torque.
In two dimensions, measuring the torques due to both a uniform expansion and an area-preserving shear of the inclusion would allow one to determine 
the three independent components of the odd viscosity tensor $\eta^o_{ijkl}$ present in a fluid with conserved angular momentum.

\section{\label{conc} Conclusions}

In the design of active materials with tailored mechanical characteristics, a basic question is:
what is the relationship between activity and mechanical response?
Whereas fluids that break both parity and time-reversal symmetries
can generically exhibit an anomalous response called odd viscosity, it remains a challenge to determine the value of this mechanical property.
When inertial effects dominate, odd viscosity is related to the
angular momentum density $\ell$ via $\eta^o = \ell/2$~\cite{Banerjee2017}.
In thermal plasmas, odd viscosity is proportional to temperature~\cite{Landau10}.
We explore a different regime, in which the fluid constituents are anisotropic and the dynamics do not conserve angular momentum.
In this regime, the equilibrium stress tensor of the fluid without drive determines the effective odd viscosity
of the active fluid once the drive is turned on.
This odd viscosity is proportional to the dissipative coefficients of nemato-hydrodynamics,
but in addition depends on the angular velocity $\Omega$ of the drive.
By modulating $\Omega$ in time, we design a classical fluid with tensorial odd viscosity. 

With this work, we aim to inspire the design of metafluids in which anomalous response can be engineered to order and observed experimentally.
Whereas in mechanical metamaterials the arrangement of the constituents leads to exotic elastic response, in these metafluids the exotic hydrodynamic response arises from time modulated drive.
These phases present an array of 
unexplored physical phenomena which combine the anisotropy
of liquid crystals with the far-from-equilibrium nature of active matter.
In addition, experimental tests of anisotropic odd viscosity
could help to elucidate this exotic and unexplored property of quantum Hall fluids in a classical fluid context.
There are two distinct experimental signatures of anisotropic odd viscosity. First, unlike its isotropic counterpart, anisotropic
odd viscosity can modify the flow in the bulk of an {\it incompressible} fluid of self-rotating object by acting as a source of vorticity, see Eq.~(\ref{eq:omm}).
Second, anisotropic odd viscosity generates torques on inclusions:
isotropic odd viscosity results in torques 
on an immersed object proportional to rate of change in its area,
whereas nematic odd viscosity results in torques
due to the rate of area-preserving shear distortion of an inclusion's shape, see Fig.~2.
The conversion between torque and shape-change via such exotic fluids
may inspire soft mechanical components and devices at the microscale.

{We thank Toshikaze Kariyado, Sofia Magkiriadou, Daniel Pearce, Alexander Abanov, William Irvine, and Tom Lubensky for insightful discussions.
AS, AG and VV were primarily supported
  by the University of Chicago Materials Research Science and Engineering
  Center, which is funded by the National Science Foundation under award number
  DMR-1420709. AG was also supported by the Quantum Materials program at
LBNL, funded by the U.S. Department of Energy under Contract
No. DE-AC02-05CH11231}

\appendix

\section{\label{sec:model1} Equilibrium nonlinear hydrodynamics}
We are interested in a fundamentally nonlinear effect: how does the
rotation rate of the nematic director affect the response to velocity gradients? 
To gain insight into this question, we examine 
contributions to the viscous stress which are higher order than 
the Ericksen-Leslie theory.
Specifically, 
terms of the form $\dot{\hat{\nv}} \nabla \vv$ in the stress tensor $\sigma_{ij}$ have 
a factor of both the director-rotation and shear rates, and contribute to the effective viscosity when the director dynamics is externally prescribed and averaged over.
Furthermore, terms with an odd number of factors of the director-rotation rate $\dot{\hat{\nv}}$
average out to zero unless the director dynamics breaks time-reversal symmetry (i.e., as long the director tip rotates by a full cycle, thereby enclosing nonzero area).
We show that terms of the order $\dot{\hat{\nv}} \nabla \vv$,  $(\dot{\hat{\nv}})^3 \nabla \vv$,
and  $(\dot{\hat{\nv}})^5 \nabla \vv$ all contribute to an effective
odd viscosity when the director $\hat{\nv}$ rotates with externally prescribed dynamics,
and that only terms of order  $(\dot{\hat{\nv}})^3 \nabla \vv$ or higher contribute to the anisotropic odd viscosity.
The term $\dot{\hat{\nv}} \nabla \vv$, averaged over rotations depends only on
the average rotation rate $\langle \dot{\hat{\nv}} \rangle$ and contributes to the isotropic odd viscosity only.

We now proceed to describe the nemato-hydrodynamic theory that includes
higher-order coupling between the $Q$-tensor and the rotation $\dot{\hat{\nv}}$.
The equation for $\mathbf{v}$ is:
\begin{equation}
\rho D_t v_i =  - \nabla_i p - \nabla_j \sigma^{0}_{ij} + \nabla_j \sigma^{EL}_{ij},
\label{eq:v}
\end{equation}
where $\sigma^{0}_{ij} = - (\nabla_{i} n_k) \partial f/ \partial \nabla_{j} n_k$ is the elastic stress tensor ($f$ is the Franck free energy density), $p$ is the pressure, and $\sigma^{EL}_{ij}$ is the Ericksen-Leslie stress on which we focus~\cite{Ericksen1959, Ericksen1961,Leslie1966,Leslie1968}.

In the usual formulation of nematohydrodynamics, 
the nematic director $\hat{\nv}(\xv,t)$ is a dynamical field that obeys a separate equation of motion.
By contrast, within our model, the nematic director is completely enslaved to an external drive.
In an experiment, this could be achieved by applying an external electric or magnetic field
so strong as to overwhelm all other terms in the equation for $\hat{\nv}(\xv,t)$.
Note that we assume this field and the director to be uniform in space, i.e., $\hat{\nv}(\xv,t) = \hat{\nv}(t)$.
This in turn significantly simplifies Eq.~(\ref{eq:v}): $\sigma^{0}_{ij}$ can be neglected.

\section{\label{sec:apdis} Hydrodynamic stresses}
We now focus on the expression for (the nonlinear generalization of) the Ericksen-Leslie stress $\sigma^{EL}_{ij}$, which is the essential ingredient in our model.
There are two equivalent approaches for writing down the form for $\sigma^{EL}_{ij}$
in terms of the strain rate components $\nabla_k v_l$, the nematic director components
$n_k$, and the director time-derivative $\dot{\hat{n}}_k$.
The original linear approach due to Ericksen and Leslie~\cite{Ericksen1959, Ericksen1961,Leslie1966,Leslie1968}
and subsequent nonlinear generalizations~\cite{Moritz1976}
include all terms allowed by symmetry, up to a given order corresponding
to the number of (hydrodynamically small) factors of $\dot{\hat{n}}_k$ and $\nabla_k v_l$ (but any number of factors of the unit vector $n_k$).
This approach has the advantage of finding all terms in a single step.
However, the approach lumps together two physically distinct contributions to $\sigma^{EL}_{ij}$:
(1) anisotropic dissipative contributions to viscous stress due to strain rate $\nabla_k v_l$ which takes into account the director $\hat{\nv}$ and (2) reactive contributions to the stress due to the nematic dynamics described by 
$\dot{\hat{\nv}}$.

The approach of, e.g., Refs.~\cite{Forster1971,Stark2003, Stark2005},
separates these dissipative and reactive contributions.
The dissipative contributions are constructed using an approach parallel to that of Ericksen and Leslie:
all terms consistent with symmetries are written down to a given order in $A_{kl} \equiv (\nabla_k v_l + \nabla_l v_k)/2$ (but not $\dot{\hat{n}}_k$).
The difference lies in the approach to
reactive terms stemming from variation of the nematic Franck free energy $F[\hat{\nv}(\xv)]$, see 
Refs.~\cite{deGennes1995,Chandrasekhar1992}.
These contributions enter the stress $\sigma^{EL}_{ij}$ via the term $\lambda_{kij} \delta F /\delta n_k$.
To make connection with the approach of Ericksen and Leslie,
we review how these reactive terms can be rewritten 
in terms of the nematic director dynamics $\dot{\hat{n}}_k$.
To do so, we use the equation of motion for the director (and include higher-order, nonhydrodynamic contributions). This nonlinear generalization of the Oseen equation reads
\begin{equation}
\frac{D n_i}{D t} = \lambda_{ijk} A_{kj} + O(A^2, A \dot{\hat{\nv}}, [\dot{\hat{\nv}}]^2) - \frac{1}{\gamma} \frac{\delta F}{\delta n_i},
\label{eq:oseen}
\end{equation}
where $D/D t$ is the material derivative of $n_i$.
The Oseen equation~(\ref{eq:oseen}) can be solved for $\delta F/\delta n_i$, and
the result substituted into $\sigma^{EL}_{ij}$.
Note that this substitution can lead to corrections of the terms in $\sigma^{EL}_{ij}$ which are nonlinear in $A$.
More significantly, these reactive terms result in all of the dependence
of  $\sigma^{EL}_{ij}$ on $\dot{\hat{\nv}}$, including terms $O(\dot{\hat{\nv}})$, $O([\dot{\hat{\nv}}]^2)$, $O(A \dot{\hat{\nv}})$,
and higher order generalizations.
This approach highlights the fact that all stresses that depend on the 
director dynamics (i.e., $\dot{\hat{\nv}}$) must ultimately arise from reactive cross-talk
between the director and the flow.
The extra step of using the Oseen equation has the advantage of providing physical intuition for
the origin of the various terms in $\sigma^{EL}_{ij}$.
However, the forms of both the linear Ericksen-Leslie terms and their nonlinear generalizations
are identical whichever approach is used to construct $\sigma^{EL}_{ij}$.

The expression for $\sigma^{EL}_{ij}$, to lowest nonlinear order~\cite{Moritz1976}, reads
\begin{widetext}
\begin{align}
\label{eq:sig}
\sigma^{EL}_{ij} &= \alpha_1 [ijkp] A_{kp} + \alpha_2 [i] N_j
+ \alpha_3[j] N_i + \alpha_4 A_{ij} + \alpha_5[ip]A_{jp}
+ \alpha_6[jp] A_{ip} + \\
&\xi_1[ijpqrs]A_{pq} A_{rs} + \xi_2 [ipqr] A_{jp} A_{qr} + \xi_3[jpqr]A_{ip}A_{qr} +\xi_4 A_{pq} A_{ij}
+ \xi_5 [ij] A_{pq} A_{pq} + \nonumber \\
& \xi_7 [pq] A_{ip} A_{jq} + \xi_8 A_{ip} A_{jp} + 
\xi_9 N_i N_j + \xi_{10} [p] A_{ip} N_j 
+ \xi_{11} [p] A_{jp} N_i + \nonumber\\
& (\xi_{12} N_p + \xi_{13} [q] A_{pq}) [i] A_{jp}
+ (\xi_{14} N_p + \xi_{15} [q] A_{pq}) [j] A_{ip}
+ \xi_{16} [ipq]A_{pq}N_j + \xi_{17} [jpq] A_{pq} N_j \nonumber
\end{align}
\end{widetext}
where $\alpha_n$ ($n = 1,\ldots,6$) are the linear nematohydrodynamic
Leslie-Ericksen coefficients, $\xi_m$ ($m = 1,\ldots,17$)
are the next-lowest-order nonlinear nematohydrodynamic coefficients ($\xi_m = \xi^1_m$ from the main text),
$N_i \equiv \dot{n}_i - W_{ij}n_j = - (\dot{\theta} - \omega) \epsilon_{ij} n_j$ is the rotation of the nematic director relative to the fluid,
and $W_{ij} \equiv \frac{1}{2} (\nabla_i v_j - \nabla_j v_i) = \omega^\prime \epsilon_{ij}$ is the antisymmetric component of the strain-rate tensor (note the difference of factor of $1/2$ between $\omega$ and $\omega^\prime$).
For outer products of the nematic director with itself, we have adopted from Ref.~\cite{Moritz1976} the notation $[ijk\cdots] = n_i n_j n_k \cdots$.

Note that in equilibrium, terms $\xi_m$ with $m = \{1,\ldots,5,16,17\}$
can be thought of as renormalizing the Leslie-Ericksen coefficients. However, in 
the calculation we consider some of these terms play distinct and important roles.
In equilibrium, the viscosity tensor $\eta_{ijkl}$ is strictly symmetric. This $4\times4$ matrix can be expressed in analogy with expression Eq.~(\ref{eq:as}):
\begin{equation}
\includegraphics{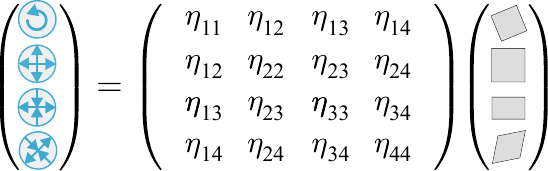}.
\label{eq:sym}
\end{equation}
The Ericksen-Leslie terms $\alpha_n$ can be re-expressed in terms of the shear viscosities and the coupling between shear and anti-symmetric stress. Note, however, that these do not include separate contributions for the isotropic bulk viscosity $\eta_{22}$. By counting the independent components, we can conclude that all of the otherviscosity terms are represented by the Ericksen-Leslie coefficients. These are the (i) shear viscosity $\eta_{33} + \eta_{44}$, (ii) amplitude $\eta_{34}^2 + (\eta_{33} - \eta_{44})^2/4$ of the anisotropic shear-shear coupling (forming the symmetric traceless component of the lower-right $2\times2$ block in Eq.~\ref{eq:sym}), (iii) amplitude $\eta_{23}^2 + \eta_{24}^2$ of coupling shear rate to isotropic stress, (iv) amplitude $\eta_{13}^2 + \eta_{14}^2$ of coupling shear rate to antisymmetric stress, (v) $\eta_{11}$ coupling of vorticity to antisymmetric stress, and (vi) $\eta_{12}$ coupling of vorticity to isotropic stress.
In equilibrium, an Onsager reciprocity relation $\alpha_6-\alpha_5=\alpha_2+\alpha_3$ further reduces these six viscosities to five independent coefficients~\cite{Parodi1970}.

\begin{widetext}
\section{\label{subsec:odd} Time averages}
The time-average of 
 a quantity $\dot{X} Y$ having one time-derivative depends only on the average rotation rate $\Omega$:
\begin{equation}
\langle \dot{X} Y \rangle = \frac{\Omega}{2 \pi} \int_0^{2 \pi/\Omega} dt \frac{d X}{d t}Y = \frac{\Omega}{2 \pi} \int_{X(0)}^{X(2 \pi/\Omega)} Y d X.
\end{equation}

We compute time-averaged expressions using
\begin{align}
 \langle n_i \dot{n}_j\rangle & = \frac{\Omega}{2 \pi} \int_0^{2 \pi/\Omega} dt \, n_i(t) \dot{n}_j(t) = 
\frac{\Omega}{2 \pi} \int_0^{2 \pi} d \theta \, n_i(\theta) n_m(\theta) \epsilon_{mj} 
= \frac{\Omega}{2} 
\epsilon_{ij}\\ 
\langle n_i \dot{n}_j n_k n_l \rangle & = \frac{\Omega}{2 \pi} \int_0^{2 \pi/\Omega} dt \, n_i(t) \dot{n}_j(t) n_k(t) n_l(t) = 
\frac{\Omega}{2 \pi} \int_0^{2 \pi} d \theta \, n_i(\theta) n_m(\theta) n_k(\theta) n_l(\theta) \epsilon_{mj}  \nonumber \\
& = - \frac{\Omega}{16} 
(\epsilon_{ik}\delta_{jl} + \epsilon_{il}\delta_{jk}  + 
 \epsilon_{jk}\delta_{il} +  \epsilon_{jl}\delta_{ik}
 - 4 \epsilon_{ij}\delta_{kl}) \equiv - \frac{\Omega}{16} (\tau_{ijkl} - 4 \epsilon_{ij}\delta_{kl}).
\end{align}
The last expression can be checked term-by-term.
These expressions differentiate the chiral active fluid from thermal averages in an isotropic equilibrium fluid:
in equilibrium fluids, there is no average rotation,
and these expressions would be zero.

We proceed by evaluating $\langle \sigma^{EL}_{ij} \rangle$ using these expressions and find
\begin{align}
\label{eq:sigav}
\langle \sigma^{EL}_{ij} \rangle &= \left( \alpha_1 \frac{A_{kk}}{4} + \frac{1}{2} \xi_{9} [\Omega-\omega^\prime]^2 \right) \delta_{ij}
+ \frac{1}{4} \left(2[\alpha_3 - \alpha_2] + [\xi_{16} -\xi_{17}] A_{kk}\right) [\Omega - \omega^\prime] \epsilon_{ij} \nonumber\\
&+ \frac{1}{2} (2 \alpha_4 + \alpha_5 + \alpha_6) A_{ij}
+ \frac{1}{2} (2 \xi_7 + \xi_8 + \xi_{13} + \xi_{15}) A_{ip}A_{jp} \nonumber\\
&+(\xi_1 \chi_{ijpqrs} + \xi_2 \phi_{ipqr} \delta_{js} + \xi_3 \phi_{jpqr} \delta_{is} + \xi_4 \delta_{ir} \delta_{js}+\frac{\xi_5}{2} \delta_{ij} \delta_{pr} \delta_{js}) A_{pq}A_{rs} 
\nonumber \\
&- \frac{\Omega - \omega^\prime}{2}
\left( \xi_{10} \epsilon_{jk} \delta_{il} A_{kl}
+\xi_{11} \epsilon_{ik} \delta_{jl} A_{kl}
- \xi_{12} \epsilon_{il} \delta_{jk} A_{kl}
- \xi_{14} \epsilon_{jl} \delta_{ik} A_{kl} \right) \nonumber \\
&- \frac{\Omega - \omega^\prime}{16} (\xi_{16} +\xi_{17})\tau_{ijpq}A_{pq}. 
\end{align}
From $\langle \sigma^{EL}_{ij} \rangle$, we can read off the form of $\eta^o$ in Eq.~[10] of the main text.
\end{widetext}

\section{Expressions for odd viscosity}
In the average stress tensor in Eq.~(\ref{eq:sigav}), the different
odd viscosity components have different prefactors $\xi_\kappa$. However, once
the forces $\nabla_j \langle \sigma^{EL}_{ij}\rangle$ are calculated in the equation for the flow $\vv$,
only a single odd viscosity term remains (of the form $\eta^o \nabla^2 \vv^*$, where $\eta^o$ is a constant).
This term has a prefactor of odd viscosity that can be read off from Eq.~(\ref{eq:sigav}) as:
\begin{equation}
\label{eq:etao2}
\eta^o = - \frac{\Omega}{8} \xi^1_L,
\end{equation}
where $\xi^\beta_L \equiv 2 [\xi^\beta_{10} + \xi^\beta_{11} - \xi^\beta_{12} - \xi^\beta_{14}] 
+ \xi^\beta_{16} +\xi^\beta_{17}$ is a linear combination of the $\xi^\beta_\kappa$ coeffiencients.
Whereas the isotropic terms from the lowest-order nonlinearities $\sigma^{EL}_{ij}$ result in the expression $\eta^o_{ij} = \eta^o \delta_{ij}$,
where $\eta^o$ is given by Eq.~(\ref{eq:etao2}), the terms from higher-order nonlinearities such as $\langle \sigma^{EL,2}_{ij} \rangle$ in the main text have contributions with magnitude
\begin{equation}
\eta^Q = \frac{\alpha \Omega^3}{4} (\xi^2_{16} + \xi^2_{17}) + O(\Omega^5).
\end{equation}
to $O(\alpha^3)$.

To obtain the expressions for components $\eta^{A}$ and $\eta^{Q}_{\gamma,\delta}$, we consider the $\omega$-dependent stress and the anti-symmetric component of the stress $\epsilon_{ij} \sigma^{EL}_{ij}/2$.
This results in the expression
\begin{equation}
    \eta^{A} = \frac{\Omega}{4}(- \xi^1_{9} + \xi^1_{10} - \xi^1_{14}) + O(\Omega^3).
\end{equation}

The anisotropic component is again higher-order in the rotation rate $\Omega$:
\begin{equation}
    \eta^{K} = \frac{\alpha \Omega^3}{4} (2 \xi^2_{11} + 2 \xi^2_{12} - \xi^2_{16} + \xi^2_{17}) + O(\Omega^5).
\end{equation}

\section{Derivation of the equation of motion}
\label{sec:derom}
Starting from the velocity equation, $\rho D_t v_j = \partial_i \sigma_{ij}$,
we substitute the stress $\sigma_{ij} = \eta_{ijkl} v_{kl}$ to arrive at 
\begin{equation}
    \rho D_t v_j = -\partial_j p + \eta \nabla^2 v_j + \partial_i \eta^o_{ijkl} v_{kl}
\end{equation}
where the first terms come from the usual treatment of pressure $p$ and dissipative isotropic shear viscosity $\eta$ and where $\eta^o_{ijkl}$ is the tensor in Eq.~(\ref{eq:as}). Defining the two components of the shear strain rate as $s^{\chi} \equiv \sigma^x_{jk} \partial_{j} v_k$ and $s^{\zeta} \equiv \sigma^z_{jk} \partial_{j} v_k$, we express the equation of motion as
\begin{widetext}
\begin{align}
    &\rho D_t v_j = \partial_j ( - p - \eta^A \omega -\eta^Q_{\alpha} s^\zeta + \eta^Q_{\beta} s^\chi) + \eta \nabla^2 v_j + \eta^o \nabla^2 (\epsilon_{jk} v_k) + \\& +\epsilon_{jk} \partial_k (\eta^A \nabla \cdot \vv + \eta^Q_{\delta} s^\chi - \eta^Q_{\gamma} s^\zeta) 
    + \sigma^z_{jk} \partial_k (\eta^Q_{\gamma} \omega  + \eta^o s^\chi + \eta^Q_{\alpha} \nabla \cdot \vv) 
    + \sigma^x_{jk} \partial_k ( - \eta^Q_{\delta} \omega - \eta^o s^\zeta  - \eta^Q_{\beta} \nabla \cdot \vv), \nonumber
\end{align}
where the first term corresponds to the pressure (i.e., the trace of the stress tensor) and vanishes in the vorticity equation. Taking the curl, the skew-gradient becomes the Laplacian: $ \epsilon_{jl} \partial_l \epsilon_{jk} \partial_k = \delta_{kl}  \partial_k \partial_l = \nabla^2$. This results in Eq.~(\ref{eq:omm2}):
\begin{align}
&\rho D_t \omega  =\eta \nabla^2 \omega - (\nabla \cdot \cM_1 \cdot \nabla) \omega + \nabla^2 [\nabla \cdot (\cM_1^* \cdot \vv)] \\&
+ (\eta^o + \eta^A) \nabla^2 (\nabla \cdot \vv) 
-  (\nabla \cdot {\mathcal M}_2 \cdot \nabla) (\nabla \cdot \vv) \nonumber ,
\end{align}
where $D_t$ is the convective derivative, and
\begin{align}
\cM_1 & \equiv \eta^Q_{\gamma} \sigma^x + \eta^Q_{\delta} \sigma^z, \\
\cM_2 & \equiv \eta^Q_{\alpha} \sigma^x + \eta^Q_{\beta} \sigma^z \nonumber
\end{align}
and $\cM_1^* \equiv \eta^Q_{\delta} \sigma^x - \eta^Q_{\gamma} \sigma^z$ (i.e., $\cM_1$ rotated by $\pi/4$).
\end{widetext}

\end{document}